\documentclass[12pt]{article}
\usepackage{amssymb}
\usepackage{amsmath}
\usepackage{epsfig}
\begin{document}

\title{Nonlinear energy transmission in the gap}
\author{J-G. Caputo, J. Leon, A. Spire\\
Physique Math\'ematique et Th\'eorique, CNRS-UMR5825\\
34095 MONTPELLIER Cedex 05 (FRANCE)}
\date{}
\maketitle

\begin{abstract}Numerical simulations of the scattering of a linear plane wave
incoming onto a nonlinear medium (sine-Gordon) reveals that: i) nonlinearity
allows energy transmission in the forbidden band, ii) this nonlinear
transmission occurs beyond an energy threshold of the incoming wave, iii) the
process begins (at the threshold) with large amplitude breathers, and then
energy is generically transmitted both by  kink-antikink pairs and breathers. 
\end{abstract}

\paragraph*{Introduction}

 From the Fermi-Pasta-Ulam recursion phenomenon \cite{fpu},
nonlinear lattice and their numerical simulations have played a key role in the
development of nonlinear studies. Since then a whole branch of theoretical
physics has evolved with the birth of the concepts of soliton \cite{zabusky}
and integrability \cite{kruskal}, for continuous, discrete and higher
dimensional systems \cite{books-ist}.  These discoveries have had a enormous
impact on the understanding of nonlinear physics \cite{books-phys}.

One of the fundamental studied problem is the energy localization (and
transport) by generation (and propagation) of nonlinear coherent structures
(solitons, breathers, nonlinear modes) by many different processes such as
external forcing, noise, local potential, inhomogeneous forcing, anharmonic
perturbation, {\em modulated} external forcing etc...  \cite{solitons}. We note
that nowadays this problem is more largely studied in the context of discrete
systems where intrinsic nonlinear modes (breathers for short) have been shown
to play the main part in storage and transport of energy \cite{discrete}.

Another basic question is the scattering of waves incoming onto a nonlinear
medium. Two classes of problems related to wave scattering have been considered
in detail: the {\em resonant wave coupling} which includes self-induced
transparency (SIT) \cite{machan}, stimulated Raman scattering
\cite{SRS}\cite{sasha}, two-photon propagation \cite{TPP}, etc..., and the {\em
self modulation} resulting from Kerr effect in optical media \cite{agrawal}
\cite{mills-book}. Within these two classes of problems, solitons have been
shown to play the central role in propagating energy in regions of parameter
(frequency, energy,...) usually forbidden by the linear theory.

The resonant wave coupling describes the process of interaction of an
electromagnetic radiation with a two-level medium through Brillouin selection
rules (the nonlinearity is extrinsic). For SIT the linear theory would conclude
with resonant absorption of radiation while nonlinearity allows for perfect
transparency. The model equations read as coupled Maxwell-Bloch system and the
scattering of waves results in a boundary-value problem for the reduced system.
The point there is that this boundary-value problem is integrable - see also
\cite{gabzak} - and it has been shown that the soliton (actually a kink for the
phase) is the  means which propagates energy through the medium without loss or
distortion, provided the input pulse obeys the {\em area theorem}
\cite{machan}. Hence in SIT, the soliton is created by a {\em sufficiently
energetic} boundary condition. In the inverse spectral transform scheme, this
boundary value problem actually maps to a Cauchy problem and the area theorem
simply states that the initial datum must produces at least one discrete
eigenvalue \cite{machan}\cite{gabzak}.

In the case of self modulation, the nonlinearity finds its origin in the medium
itself (intrinsic Kerr effect). A process similar to SIT (corresponding to a
Cauchy problem in the IST scheme), but with different origin, occurs in fiber
guides in nonlinear regime (and with anomalous dispersion) where a sufficiently
energetic input light pulse  evolves as a soliton of the nonlinear
Schr\"odinger equation (NLS) along the fiber \cite{has-moll} \cite{agrawal}.
Here and in SIT, the energy flows with solitons created by a localized high
energy pulse.

A different soliton effect arises in the study of the propagation of light in a
medium with forbidden bands (photonic band gaps) for which the concept of {\em
gap soliton} have been introduced \cite{mills}.  There it has been shown that
the nonlinearity in a dielectric periodic thin-film stack (medium with spatial
periodic variation of the refraction index) allows for transmission of
radiation inside the gap by means of an envelope soliton. In short, the
presence of an envelope soliton in the medium shifts locally the gap and allows
for transparency \cite{lilisou}. The model for the slowly varying envelope is
the NLS equation \cite{sipe}.  For that reason (transmission of radiation in
the gap), and the related switching properties (tranmission with soliton -
reflection without), photonic band gap media are the subject of active
research, see e.g.  \cite{soukoulis}. Notice that the {\em creation} of the
envelope soliton in those media is an open question. Indeed, while the
nonlinearity, via NLS, offers possible existence of a soliton, the mechanism
of its generation itself is still to be understood. Such a mechanism has been
for instance analytically described in the case of stimulated Raman scattering
in \cite{sasha} with the tool of the inverse spectral transform applied to
boundary value problems (another example can be found in \cite{fx}).

Then the mechanism of soliton creation by radiation is of fundamental interest
and we consider here the (nonlinear) scattering of waves from a quite
elementary point of view. Namely we  explore the possibility for a nonlinear
medium to build up nonlinear coherent structures under {\em ``exposition''} to
input radiation.  The problem we are interested in, very simple in linear
systems, concerns the reflection and transmission of a plane wave in an
inhomogeneous medium with band pass type dispersion relation: a plane wave
created in a medium 1 is sent onto a medium 2 with a different dispersion
relation.  In particular total reflection occurs when the incoming wave of
medium 1 falls onto medium 2 with a frequency inside a frobidden band.  We
consider then this generic question in the case when the medium 2 is
intrinsically nonlinear and modelized by an integrable equation (sine-Gordon).  

\paragraph*{Statement of the problem.}

We consider then two different media defined by the wave equations
\begin{align}
&x<0\ :\quad u_{tt}-c_1^2u_{xx}+\omega_1^2u=0\ ,\label{cont1}\\
&x>0\ :\quad u_{tt}-c_2^2u_{xx}+\omega_2^2\sin u=0\ ,\label{cont2}
\end{align}
with continuity conditions in $x=0$ (inferred here from the natural discrete
version of the above system, see later).  The scattering of a wave of given
frequency $\Omega$ and normalized amplitude incoming from medium 1 to medium 2,
when $\Omega$ lies in the phonon gap of the nonlinear medium 
($\Omega<\omega_2$), is described {\em in the linear limit} by the general  
solution (the normalization is chosen such as to have an incident wave
$A\sin(Kx-\Omega t)$)
\begin{align}
&x<0\ :\quad u(x,t)=\frac1{2i}e^{-i\Omega t}\left[Ae^{iKx}+Be^{-iKx}\right]+
{\rm c.c.}\ , \label{solmed1}\\
&x>0\ :\quad u(x,t)=\frac1{2i}e^{-i\Omega t}\left[Ce^{-\kappa x}+De^{\kappa x}
\right]+{\rm c.c.}\ , \label{solmed2} \end{align}
with  $K$ and $\kappa$ given from $\Omega$ by
\begin{equation}\label{disp}
\Omega^2 =\omega_1^2+K^2c_1^2 = \omega_2^2-\kappa^2c_2^2\ .
\end{equation}

While the linear theory produces an evanescent wave ($D=0$), the nonlinearity
does not prevent in principle nonlinear superposition\footnote{See the comment
at the end of the paper}  of exponentials in \eqref{solmed2}, possibly allowing
energy transmission  to the nonlinear medium by means of a soliton.  We address
then the question wether or not solitons build up under scattering and 
allow nonlinear transmission of radiation within the gap. 

The present study is fully based on numerical simulations and, before going
further, it is worth stressing that we will discover that indeed solitons
(actually kink-antikink pairs and breathers) are created to allow for energy
transmission. Moreover, this process is in essence different either from
self-induced transparency in two-level media, or else from gap solitons in
photonic band gap media, because reductive perturbation method (or multi-scale
expansions) do not apply: the incoming radiation creates {\em large amplitude}
nonlinear objects.

\paragraph*{Method.}

The system \eqref{cont1}\eqref{cont2} is first maped to the following discrete
analog where the first nonlinear pendulum $u_1$ is coupled to the
last linear pendulum $u_0$ by the spring $c_1$ of the linear chain:
\begin{align}\label{disc-sys}
n<1\ :\quad& 
\ddot u_n-c_1^2(u_{n+1}-2u_n+u_{n-1})/h^2+\omega_1^2u_n=0\ , \notag\\
n=1\ :\quad& 
\ddot u_1-c_2^2(u_2-u_1)/h^2+c_1^2(u_1-u_0)/h^2+\omega_2^2\sin u_1=0\ ,\notag\\
n>1\ :\quad& 
\ddot u_n-c_2^2(u_{n+1}-2u_n+u_{n-1})/h^2+\omega_2^2\sin u_n=0\ .
\end{align}
This maps to \eqref{cont1}\eqref{cont2} in the continuous limit $h\to0$ and
has natural unconstrained continuity conditions in $n=1$.

\begin{figure}[ht]
\centerline{\epsfig{file=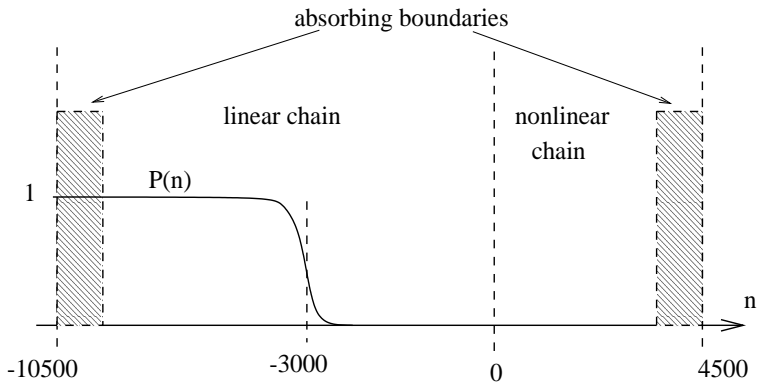,height=5cm,width=10cm}}
\caption{\em Grid data for numerical integration}\label{fig:chain}
\end{figure}

In order to generate in the linear medium 1 an {\em incoming plane wave} we
follow the work \cite{jean-guy} by making the change of variable from $u_n(t)$
to $f_n(t)$ :
\begin{equation}\label{wave}
u_n(t)=f_n(t)+AP(n)\ \sin(Knh-\Omega t)\ ,\end{equation}
where $\Omega$ is given from $K$ by \eqref{disp}, where $A$ is the amplitude of
the incoming wave and where the fuction $P(n)$ (a polynomial) is chosen such as
to vary from the value $1$ for large negative $n$ to the value $0$ far before
the nonlinear medium 2.  To set things clear, we have used a medium 1 made of a
(linear) chain of $10500$ oscillators and the plynomial $P(n)$ varies from the
value 1 at $n=-10500$ to zero around $n=-3000$. The medium 2 (nonlinear chain)
is constituted of $4500$ pendula.  Finally, to simulate the infinite line,
strong damping is included at both ends of the interval (so called {\em
absorbing boundaries}) on a lengh of about $450$ particles. All these settings
are summarized in figure \ref{fig:chain}.  
The link with the continuous system
is acheived with step $h=1/30$, for which we shall check that the generated
nonlinear structures are continuous objects.

\begin{figure}[ht]
\centerline{\epsfig{file=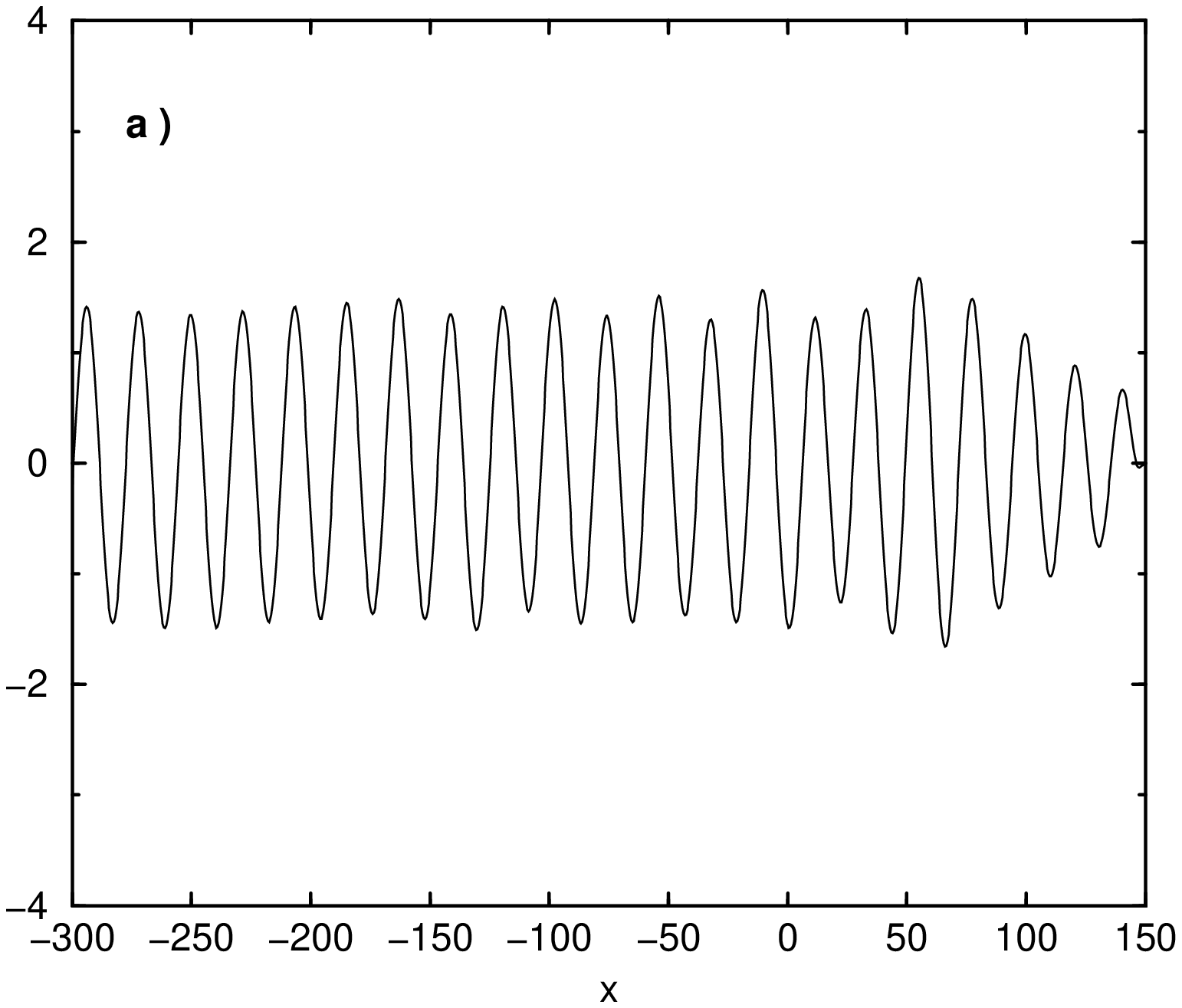,height=5cm,width=6cm}
\epsfig{file=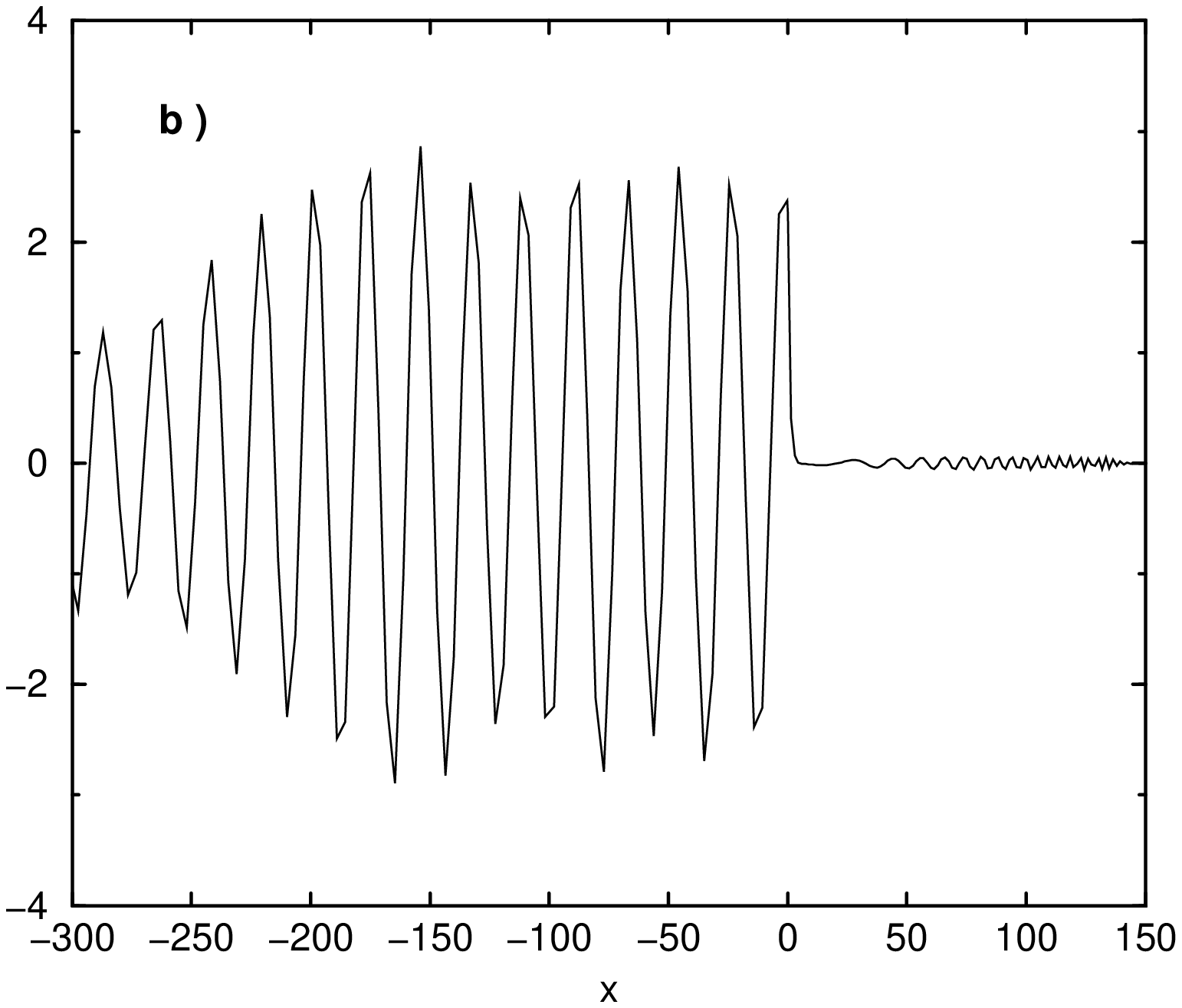,height=5cm,width=6cm}}
\caption{\em Graphs of $u(x,t)$ at $t=250$.  
a) evidence of plane wave generation in a linear homogenous medium. 
b) evidence of total reflection on the stop gap
of a linear right-hand-side medium. }\label{linlin}\end{figure}

The expression \eqref{wave} is inserted in system \eqref{disc-sys} which is
then solved for $f_n(t)$ as a system of $2N$ coupled first order differential
equations in the variable $t$ ($N=15000$) with the subroutine DOPRI 5. This
method allows to launch a plane wave in the system and its efficiency is first
checked by equating both media to the linear one and we obtain a nice plane
wave propagating without distortion for very long times (longer than the actual
nonlinear experiments) as  seen on figure \ref{linlin}(a) where the  parameters
of the (homogeneous, linear) medium are $c_1^2=8,\ \omega_1^2=1$ and the input
wave has frequency $\Omega=1.29$ and amplitude $A=1.42$.  Note in particular
that the created wave does vanish for large $x=nh$ thus testing the efficiency
of the absorbing boundaries (the left-hand side absorbing boundary from 
$x=-300$ to $x=-315$ is not represented but works as well).

Second we have verified that, if the right hand side medium is linear, a wave
with frequency in the phonon gap is indeed totally reflected as seen on figure
of \ref{linlin}(b), in which the amplitude of the incident wave largely exceeds
the threshold amplitude required for nonlinear transmission described later.
Here the parameters are $c_1^2=8,\ \omega_1^2=1$ for the left (linear) medium
and $c_2^2=1,\ \omega_2^2=3$ for the right (still linear) medium. The frequency
of the input plane wave is $\Omega=1.29<\omega_2$ and its amplitude $A=1.5$.

\paragraph*{Results.}

Thanks to the scale invariance of the system \eqref{cont1}\eqref{cont2}, we can
always fix the parameters of one of the two media and vary the others. Herafter
we shall then chose the parameters of the nonlinear medium to be
\begin{equation}\label{param}
c_2^2=1\ ,\quad\omega_2^2=3\ ,\end{equation}
and vary the linear medium ($c_1$ and $\omega_1$) and the incident plane wave
$A\sin(Kx-\Omega t)$. We obtain the following results.

\begin{figure}[ht]
\centerline{\epsfig{file=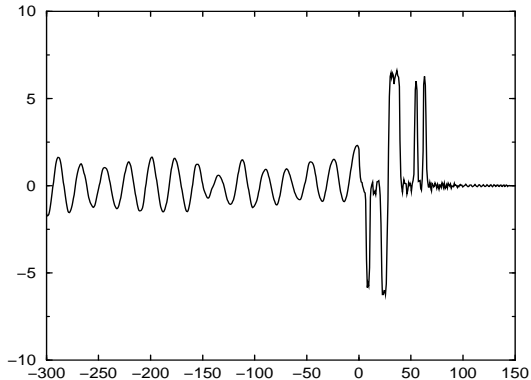,height=5cm,width=7cm}}
\caption{\em Solution $u(x,t)$ on the full line at $t=120$ for a incident wave 
at $\Omega=1.29$ and energy $E=1.8$. The parameters of the linear 
chain are $c_1^2=8,\ \omega_1^2=1$.}\label{ligne}
\end{figure}

\begin{figure}[ht]
\centerline{\epsfig{file=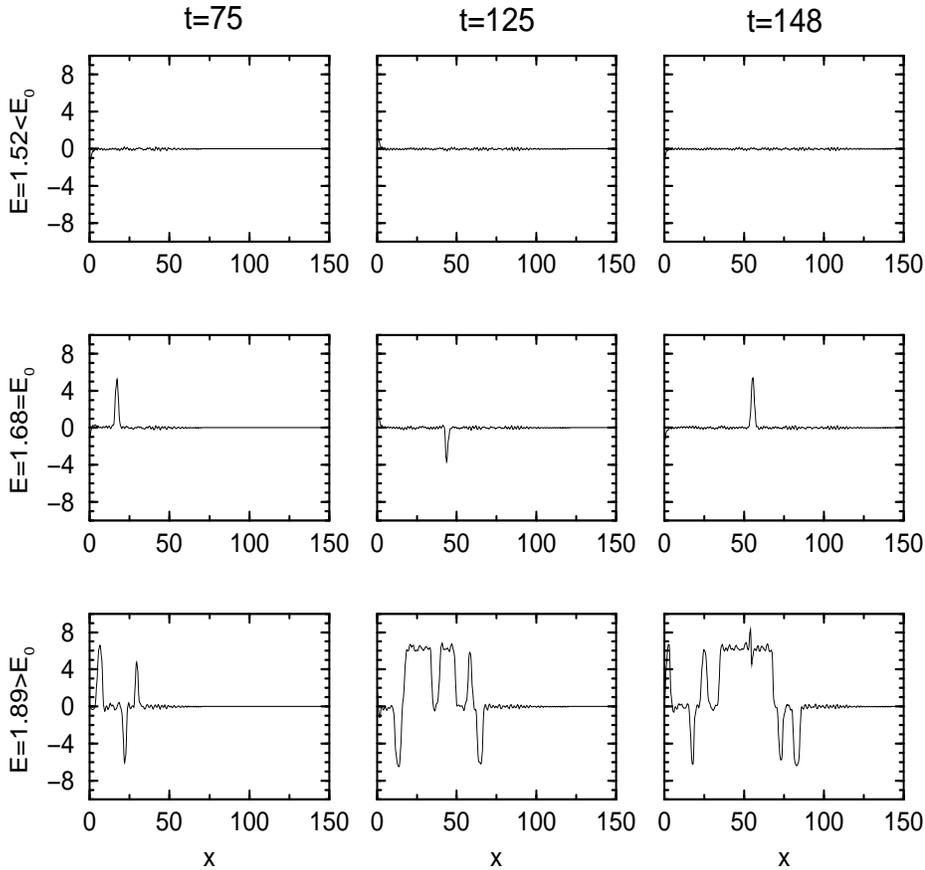,height=12cm,width=12cm}}
\caption{\em Representation of nonlinear oscillators $u(x,t)$ 
for a incident wave 
at $\Omega=1.29<\omega_2$ as a function of $x$ for different times. 
The energy $E$ is varied with the amplitude $A$
of incident wave. The parameters of the linear chain are $c_1^2=8,\ 
\omega_1^2=1$.}\label{bion}
\end{figure}

1 - An incoming wave with frequency inside the gap of the nonlinear medium
($\Omega<\omega_2$) is transmitted if its average energy $E$ exceeds a
threshold energy $E_0$. For the choice \eqref{param} we found $E_0=1.68$. The
value of $E$ we refer to is obtained by averaging over one period the
hamiltonian of the linear chain i.e.
\begin{equation}E=\frac12A^2\Omega^2\ .  \end{equation} 
This energy transmission is accomplished by the generation of kink-antikink
pairs and large amplitude breathers. The former occur alone when the incident
energy equals the threshold $E_0$.  Both  breathers and kink-antikinks then
propagate in the medium.  To illustrate this, a typical experiment at energy
$E>E_0$ is shown on figure \ref{ligne} where the function $u(x,t)$ is drawn at
time $t=120$ in both media. By looking the picture at different times, we 
can state that the transmitted wave is made of two breathers followed by two 
kink-antikink pairs, followed themselves by a new breather. 
The various behaviors are then depicted on
fig.\ref{bion} where the wave in the nonlinear medium is drawn for different
times either for an incident energy $E<E_0$ (total reflection), $E=E_0$
(breather generation) and $E>E_0$ (kink-antikink and breather creation). Note
that the average full width at half maximum of the breather, e.g. the one
appearing in fig.\ref{bion} at  $E=E_0$, contains about $73$ particles, hence 
being a continuous object.

\begin{figure}[ht]
\centerline{\epsfig{file=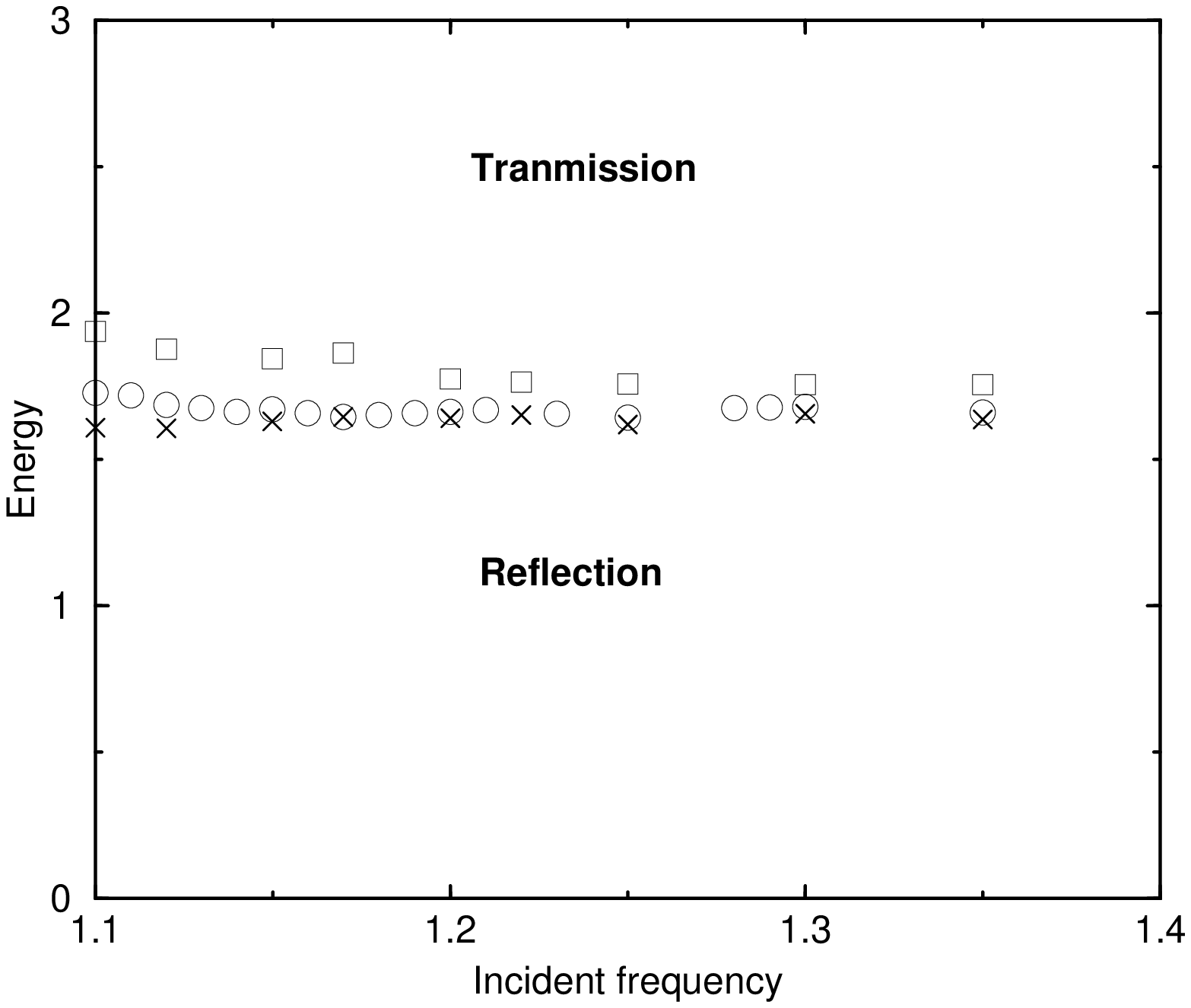,height=5cm,width=7cm}}
\caption{\em Energy threshold for nonlinear absorption. The parameters of the
linear chain are:
$\{\square\ \to\ c_1^2=10,\  \omega_1^2=0\}$, 
$\{\circ \ \to\  c_1^2=8,\  \omega_1^2=1\}$ 
$\{\times\ \to\  c_1^2=16,\  \omega_1^2=1\}$. }\label{seuil}
\end{figure}

2 - The found threshold energy $E_0$ for nonlinear transmission has the 
following properties: it does not depend on the incident 
frequency $\Omega$ and it does not depend either on the {\em nature} of 
the linear left-hand-side medium (electromagnetic vacuum for $\omega_1^2=0$
or dielectric medium for $\omega_1^2\ne0$).
This is illustrated on fig.\ref{seuil} where the dots
denote values above which nonlinear transmission becomes effective as functions
of the incident frequency $\Omega$.  

\begin{figure}[ht]
\centerline{\epsfig{file=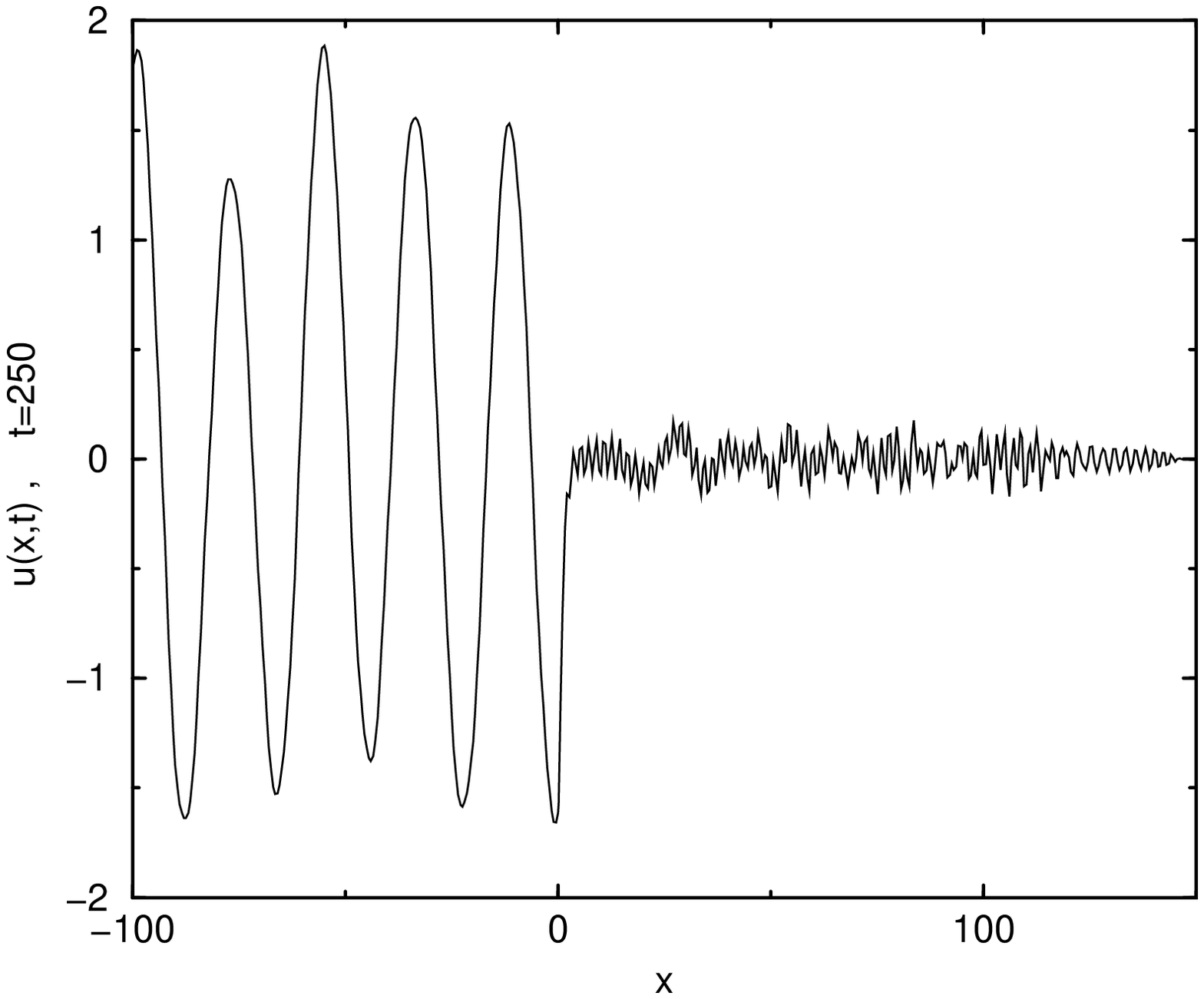,height=5cm,width=7cm}}
\caption{\em This figure shows the solution $u(x,t)$ at $t=250$ 
in the case $E=1.89>E_0=1.68$ as in fig.\ref{bion} when $\sin u$ is replaced 
with its Taylor expansion at order $u^7$.}\label{nlkg}
\end{figure}

3 - This transmission process does not occur if the nonlinear system does not
possess soliton solutions. For instance, replacing the nonlinearity $\sin u$ by
any of its truncated Taylor expansions, we do not obtain energy transmission. 
We illustrate this on fig.\ref{nlkg} where $\sin u$ has been replaced by its
Taylor expansion up to order $u^7$.  This indicates that this process does not
result from a shift of the stop gap due to nonlinearity. If such would be the
case, one would also observe transmission with the Taylor expansion of 
$\sin u$  (and moreover one would also observe transmission of radiation as 
{\em waves}, not as nonlinear objects).

\paragraph*{Conclusion}

By modelizing the scattering of a plane wave incoming onto a nonlinear medium,
we have discovered the phenomenon that we call {\em nonlinear
transmission}, and which relies on the property of a soliton-bearing system
(like sine-Gordon but unlike nonlinear Klein-Gordon) of being able to
superimpose nonlinearly exponential waves. With respect to gap solitons where
the incident radiation flows through the medium thanks to the presence of a
soliton who shifts the gap (or {\em ``opens the door''}), here the radiation is
totally converted into localized large amplitude nonlinear objects. These
nonlinear objects cannot be described by perturbation approaches but require
fully nonlinear treatment (e.g. via inverse spectral transform). 

This nonlinear transmission has immediate important implications such as {\em
nonlinear absorption} of radiation (which could play a role in the
interpretation of anomalous absorption bands), or else {\em nonlinear
tunelling}, both in continuous and discrete systems. All this issues will be
displayed in forthcoming works but we want to stress here that the theoretical
interpretation of the threshold $E_0$ of incident average energy is still to be
understood.

Note that we have checked that this threshold is not due to discreteness
effects (Peierls barrier to ovecome) by duplicating the numerical simulations
with different steps. Varying the step size from $h=1/10$ to $h=1/100$ we have
obtained the same threshold of nonlinear transmission.

\paragraph*{Comment}

The nonlinear superposition of exponentials, say $\exp[\xi_1(x,t)]$ and
$\exp[\xi_2(x,t)]$, is obtained by just writing down the two-soliton solution
of sine-Gordon under the form
\begin{equation}\label{k-kbar}
u(x,t)=4\arctan\left[\frac{a_1+a_2}{a_1-a_2}\frac{e^{\xi_2}-e^{\xi_1}}
{1+e^{\xi_2+\xi_1}}\right]\ ,\end{equation}
where $a_1$ and $a_2$ are two arbitrary parameters, and where
\begin{equation}
2\xi_j=(a_j+1/a_j)\omega_2/c_2(x-x_j)+(a_j-1/a_j)\omega_2(t-t_j)\ ,
\end{equation}
with arbitrary constants $x_j$ and $t_j$ (note that those values
allows us to fix arbitrarily  amplitudes and phases of both exponentials).
It is remarkable that the functions $\exp[\xi_j(x,t)]$ are precisely 
solutions of the linearized sine-Gordon equation
\begin{equation}
v_{tt}-c_2^2v_{xx}+\omega_2^2 v=0\ ,\end{equation}
and the formula \eqref{k-kbar} indeed furnishes a {\em nonlinear 
superposition recipe}. 

The kink-antikink results from parameters $a_j$ real and of {\em same sign}
(opposite signs produce the 2-kink or the 2-antikink solutions). The breather
is obtained by assuming complex constants $a_j$ with $a_2=\bar a_1$.  A quite
simple exercise allows one to have insight on the creation and propagation of
such nonlinear structures.  In particular the kink and antikink will be seen to
separate as time runs, as on the numerical experiments in fig.\ref{bion} (use
for instance the parameters $a_1=-0.85$, $a_2=-0.9$, $x_1=5$, $x_2=5$, $t_j=0$
in the normalized units, i.e. for $c_2=\omega_2=1$).

We note finally that the stationary breather furnishes a striking instance of a
nonlinear superposition of the evanescent waves in \eqref{solmed2}. Indeed
defining
\begin{equation}
a_1=\bar a_2=-\kappa\frac{c_2}{\omega_2}+i\frac{\Omega}{\omega_2}
\end{equation}
we obtain for $x_j=t_j=0$
\begin{equation}
u(x,t)=4\arctan\left[i\frac{\kappa c_2}{\Omega}\ 
\frac{e^{-i\Omega t-\kappa x}-e^{i\Omega t-\kappa x}}
{1+e^{-2\kappa x}}\right]\ ,\end{equation}
where the relation \eqref{disp} between $\kappa$ and $\Omega$ simply
means that we have chosen $|a_j|=1$ (for a {\em stationary} breather).

\end{document}